\documentclass[a4paper,11pt]{article}
\usepackage{pos}
\usepackage{graphicx} 
\usepackage{subcaption}
\usepackage{wrapfig}

\title{Multiwavelength study of Galactic PeVatron LHAASO J0341+5258}
 \ShortTitle{Multiwavelength study of Galactic PeVatron LHAASO J0341+5258}

\author*[a]{P. Bangale}
\author[]{for the VERITAS Collaboration}
\author[b]{X. Wang}
\onbehalf{for the HAWC Collaboration}
\affiliation[a]{Department of Physics, \\ Temple University, Philadelphia, PA 19107, USA}

\affiliation[b]{Department of Physics, \\ Michigan Technological University, Houghton, MI 49931, USA}

\emailAdd{priyadarshini.bangale@temple.edu}

\abstract{Galactic PeVatrons are astrophysical sources accelerating particles up to a few PeV (~10$^{15}$ eV). The primary method to identify both electron and proton PeVatrons is the observation of $\gamma$-ray radiation at ultra-high energies (UHE; E$>$100 TeV). In 2021, LHAASO detected 14 steady $\gamma$-ray sources with photon energies above 100 TeV and up to 1.4 PeV. Most of these sources can be plausibly associated with objects such as supernova remnants, pulsar wind nebulae, and stellar clusters. However, LHAASO J0341$+$5258 is detected as an unidentified PeVatron, emitting $\gamma$ rays at energies above hundreds of TeV. It is extended in nature and notably bright, with a flux $>$ 20\% of the Crab Nebula's flux above 25 TeV. Multiwavelength observations are required to identify the PeVatron responsible for the UHE $\gamma$ rays, understand the source morphology and association, and shed light on the emission processes. Here, we will present the results from the VERITAS and HAWC observations of this PeVatron, along with a discussion on potential emission scenarios through multiwavelength modeling.}

\ConferenceLogo{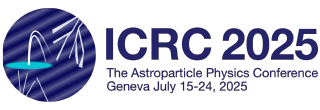}

\FullConference{39th International Cosmic Ray Conference (ICRC2025)\\
 15–24 July 2025\\
Geneva, Switzerland\\}

\begin{document}
\maketitle

\section{Introduction}
Understanding the nature and emission mechanisms of Galactic PeVatrons is a critical step toward understanding the longstanding problem of cosmic ray origin.~$\gamma$-ray observations in the ultra-high energy (UHE; E$>$100\,TeV) band have proven to be helpful probes towards the PeVatron searches \citep{Abeysekara2020, cao2021a, cao2023, cao2021b, Aharonian2021}. Investigating the TeV–PeV energy range is essential for identifying sources that can explain cosmic rays up to the knee and beyond. Therefore, the search for PeVatrons has been one of the key science topics for VERITAS and HAWC. In 2021, LHAASO detected 14 steady $\gamma$-ray sources with photon energies above 100\,TeV, up to 1.4\,PeV \citep{cao2021a, cao2021b, Aharonian2021}. This finding has provided a clear direction for the ongoing PeVatron searches with VERITAS and HAWC. \\

LHAASO J0341+5258 is an unidentified $\gamma$-ray source based on 308.33 days of data from the Kilometer Squared Array (KM2A) \citep{cao2021b}. The source is extended in nature, with size of $(0.29\pm0.06_{stat}\pm0.02_{sys})^{\circ}$, and is notably bright, exhibiting a flux greater than 20\% of that of the Crab Nebula above 25 TeV. In 2023, LHAASO published its first comprehensive catalog featuring 90 sources, including 43 UHE sources above 100 TeV, based on 508 days of Water Cherenkov Detector Array (WCDA) data and 933 days of KM2A data. In this catalog, LHAASO\,J0341+5258 was resolved into two distinct sources using KM2A: 1LHAASO\,J0339+5307 and 1LHAASO\,J0343+5254u*.~Additionally, 1LHAASO\,J0343+5254u* was detected in the 1-25 TeV energy range using the WCDA detector, with a test statistic (TS) of 94.1 and a similar extension to LHAASO\,J0341+5258~(0.33$\pm$0.05)$^{\circ}$.~Using KM2A, 1LHAASO J0339+5307 was detected at an offset of 0.37$^{\circ}$ from the position of LHAASO J0341+5258, with a TS of 144 and an extension of $<$0.22$^{\circ}$, whereas 1LHAASO J0343+5254u* was detected at an offset of 0.28$^{\circ}$, with a TS of 388.1 and an extension of (0.20$\pm$0.02)$^{\circ}$.  Note that these extension of LHAASO source regions are the 39\% containment radius of the 2D-Gaussian model (r\_39 region) reported in the LHAASO catalog Since VERITAS covers the same energy range as WCDA, VERITAS can, in principle, provide a complementary view of this source but with the added advantage of better angular and energy resolution.

Studies of this region in other wavelengths include recent X-ray observations of 120 ks conducted in February 2024 with XMM-Newton have revealed a candidate pulsar wind nebula (PWN), a possible counterpart for the LHAASO J0343+5254u \citep{DiKerby2025}. This source is extended, with an angular size of 0.03$^{\circ}$. In addition, CO observations of the LHAASO J0341+5258 region, conducted by the Milky Way Imaging Scroll Painting (MWISP) project \citep{Su2019}, reveal partially overlapping molecular gas in the form of a half-shell structure \citep{cao2021b}. The total mass of gas estimated within 1$^{\circ}$ of the LHAASO source is about 10$^3$ M$_{\odot}$, assuming a distance of 1 kpc, with no clear CO emission detected at greater distances. 

\section{Observations and results}

\begin{figure*}
    \begin{subfigure}[b]{0.45\columnwidth}
        \includegraphics[scale=0.585]{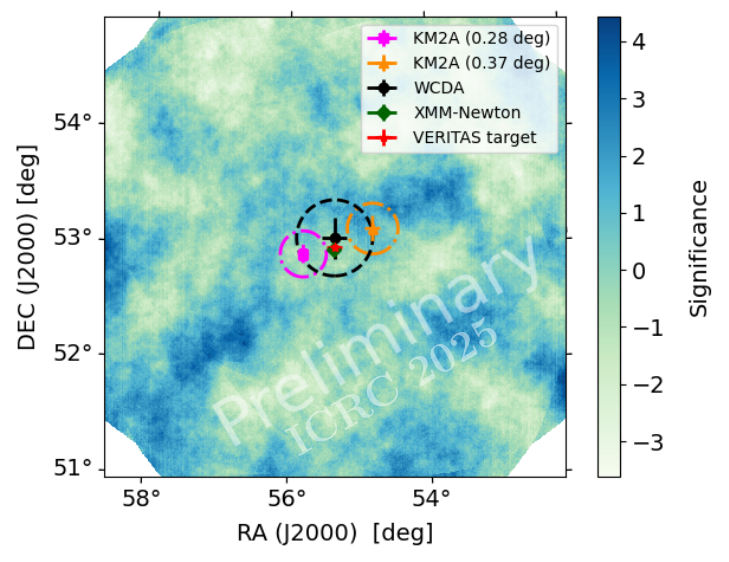}
        \caption{VERITAS}
        \label{fig:veritas_skymap}
    \end{subfigure}
    \hfill
    \begin{subfigure}[b]{0.5\columnwidth}
        \includegraphics[scale=0.58]{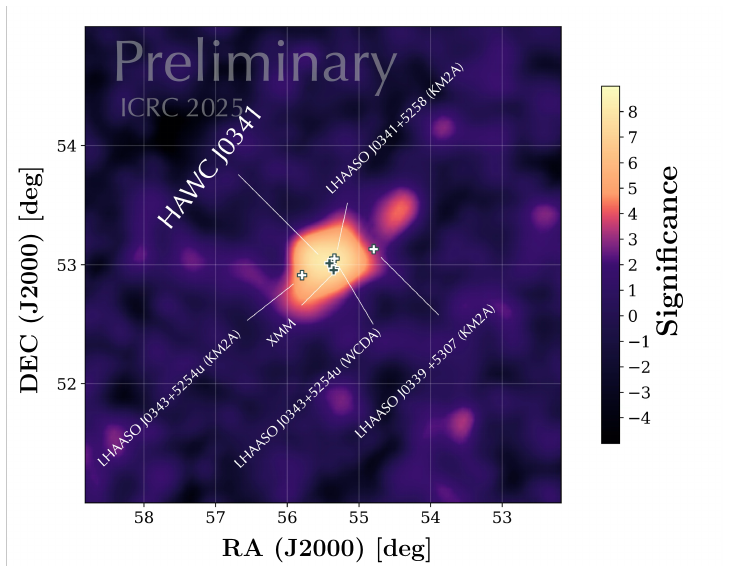}
        \caption{HAWC}
        
        \label{fig:hawc_skymap}
    \end{subfigure}
    \hfill
    \begin{subfigure}[b]{1\columnwidth}
    \end{subfigure}
    \caption{Significance maps for VERITAS (left panel) and HAWC (right panel) data in the region centered on LHAASO\,J0341+5258 position. The source is clearly detected in HAWC data, while there is no significant detection in VERITAS data. On the left plot, the WCDA extension region from 1LHAASO\,J0343+5254u* is shown in a dashed black circle. Additionally, the VERITAS significance map features two extension regions: 1LHAASO J0343+5254u* and LHAASO J0341+5307, corresponding to KM2A, with offsets of 0.28$^{\circ}$ and 0.37$^{\circ}$, respectively.  Note that these extension regions are the 39\% containment radius of the 2D-Gaussian model (r\_39 region) reported in the LHAASO catalog \citep{cao2023}. The XMM-Newton-detected PWN is shown in green solid diamond.}
    \label{fig:four figures}
\end{figure*}

To investigate the TeV counterparts, we analyzed data from VERITAS \citep{Park2015} and HAWC \citep{Abeysekara2017}. The VERITAS observations were conducted from October 2021 to January 2022 and again from September 2022 to January 2023, totaling 50 hours of data.~Observations were paused during periods when the source was not visible and during the annual Arizona monsoon season. The analyses were performed using the EventDisplay analysis package \citep{ED2017} and subsequently cross-checked with the VEGAS analysis package \citep{vegas2008}. Given the extended nature of the source from WCDA (0.33$^{\circ}$), we searched the VERITAS data using an extended source analysis, applying an integration region ($\theta$) of 0.25$^{\circ}$ centered around the position of LHAASO\,J0341+5258. No significant emission was detected in this dataset (see Figure \ref{fig:veritas_skymap}).

For the HAWC analysis, we utilized the neural network energy estimator dataset comprising 2769 days of data from June 2015 to January 2023. The neural network is one of two independent energy estimators used by HAWC. We binned the data based on the fraction of PMTs that were hit\footnote{We calculate the fraction of PMT channels as PMT\_triggered/PMT\_available for each event.} and energies, as outlined in the HAWC Crab analysis from 2019 \citep{Abeysekara2019}. We selected a circular region of interest with a radius of 3$^{\circ}$ around the position of LHAASO\,J0341+5258. The source was detected with a significance level of approximately 8.2\,$\sigma$ (refer to Figure \ref{fig:hawc_skymap}). We tested both the point source model and the extended source model using a simple power law, finding that the extended source template is preferred over the point-source template by $\Delta\rm{TS}$ of 18, with an extension calculated at (0.18$\pm$0.04)$^{\circ}$.

\section{Discussion}

Recent X-ray observations conducted with XMM-Newton have identified a new extended PWN-like source within the region of interest (ROI) of LHAASO J0341+5258 \citep{DiKerby2025}. It exhibits characteristics similar to other PWNe, such as Eel, Boomerang, and Dragonfly, where smaller X-ray regions have been associated with larger TeV regions \citep{Burgess2022, Pope2024}. This suggests that the TeV emission is likely of leptonic origin, resulting from the inverse Compton scattering of relativistic electrons within the surrounding PWN \citep{Aharonian2004, Sudoh2019}. However, there is a caveat: a powerful pulsar has not been detected in the PWN identified by XMM-Newton. This lack of detection could be due to several factors: the pulsar might be absent, its beam could be misaligned relative to the observer’s line of sight, or its spin-down luminosity may be below the threshold needed to produce detectable emission \citep{Guillemot2015}. 

Additionally, since a molecular cloud partially overlaps the region of LHAASO J0341+5258, there is an alternative explanation that involves a hadronic emission scenario. This scenario may arise from an old supernova remnant (SNR) from which cosmic rays have already escaped, resulting in interactions with a nearby molecular cloud and accumulating in this currently invisible SNR \citep{Gabici2007}. Our next step will be to conduct detailed multiwavelength spectral modeling to explore the various possible emission scenarios. 

\acknowledgments
This research is supported by grants from the U.S. Department of Energy Office of Science, the U.S. National Science Foundation and the Smithsonian Institution, by NSERC in Canada, and by the Helmholtz Association in Germany. This research used resources provided by the Open Science Grid, which is supported by the National Science Foundation and the U.S. Department of Energy's Office of Science, and resources of the National Energy Research Scientific Computing Center (NERSC), a U.S. Department of Energy Office of Science User Facility operated under Contract No. DE-AC02-05CH11231. We acknowledge the excellent work of the technical support staff at the Fred Lawrence Whipple Observatory and at the collaborating institutions in the construction and operation of the instrument. 

We acknowledge the support from: the US National Science Foundation (NSF); the US Department of Energy Office of High-Energy Physics; the Laboratory Directed Research and Development (LDRD) program of Los Alamos National Laboratory; Consejo Nacional de Ciencia y Tecnolog\'{i}a (CONACyT), M\'{e}xico, grants 271051, 232656, 260378, 179588, 254964, 258865, 243290, 132197, A1-S-46288, A1-S-22784, CF-2023-I-645, c\'{a}tedras 873, 1563, 341, 323, Red HAWC, M\'{e}xico; DGAPA-UNAM grants IG101323, IN111716-3, IN111419, IA102019, IN106521, IN110621, IN110521 , IN102223; VIEP-BUAP; PIFI 2012, 2013, PROFOCIE 2014, 2015; the University of Wisconsin Alumni Research Foundation; the Institute of Geophysics, Planetary Physics, and Signatures at Los Alamos National Laboratory; Polish Science Centre grant, DEC-2017/27/B/ST9/02272; Coordinaci\'{o}n de la Investigaci\'{o}n Cient\'{i}fica de la Universidad Michoacana; Royal Society - Newton Advanced Fellowship 180385; Generalitat Valenciana, grant CIDEGENT/2018/034; The Program Management Unit for Human Resources \& Institutional Development, Research and Innovation, NXPO (grant number B16F630069); Coordinaci\'{o}n General Acad\'{e}mica e Innovaci\'{o}n (CGAI-UdeG), PRODEP-SEP UDG-CA-499; Institute of Cosmic Ray Research (ICRR), University of Tokyo. H.F. acknowledges support by NASA under award number 80GSFC21M0002. We also acknowledge the significant contributions over many years of Stefan Westerhoff, Gaurang Yodh and Arnulfo Zepeda Dominguez, all deceased members of the HAWC collaboration. Thanks to Scott Delay, Luciano D\'{i}az and Eduardo Murrieta for technical support.


\clearpage

\begin{center} 
\section*{\centering{All Authors and Affiliations}} 
\end{center}

\section*{Full Author List: VERITAS Collaboration}

\scriptsize
\noindent
A.~Archer$^{1}$,
P.~Bangale$^{2}$,
J.~T.~Bartkoske$^{3}$,
W.~Benbow$^{4}$,
Y.~Chen$^{5}$,
J.~L.~Christiansen$^{6}$,
A.~J.~Chromey$^{4}$,
A.~Duerr$^{3}$,
M.~Errando$^{7}$,
M.~Escobar~Godoy$^{8}$,
J.~Escudero Pedrosa$^{4}$,
Q.~Feng$^{3}$,
S.~Filbert$^{3}$,
L.~Fortson$^{9}$,
A.~Furniss$^{8}$,
W.~Hanlon$^{4}$,
O.~Hervet$^{8}$,
C.~E.~Hinrichs$^{4,10}$,
J.~Holder$^{11}$,
T.~B.~Humensky$^{12,13}$,
M.~Iskakova$^{7}$,
W.~Jin$^{5}$,
M.~N.~Johnson$^{8}$,
E.~Joshi$^{14}$,
M.~Kertzman$^{1}$,
M.~Kherlakian$^{15}$,
D.~Kieda$^{3}$,
T.~K.~Kleiner$^{14}$,
N.~Korzoun$^{11}$,
S.~Kumar$^{12}$,
M.~J.~Lang$^{16}$,
M.~Lundy$^{17}$,
G.~Maier$^{14}$,
C.~E~McGrath$^{18}$,
P.~Moriarty$^{16}$,
R.~Mukherjee$^{19}$,
W.~Ning$^{5}$,
R.~A.~Ong$^{5}$,
A.~Pandey$^{3}$,
M.~Pohl$^{20,14}$,
E.~Pueschel$^{15}$,
J.~Quinn$^{18}$,
P.~L.~Rabinowitz$^{7}$,
K.~Ragan$^{17}$,
P.~T.~Reynolds$^{21}$,
D.~Ribeiro$^{9}$,
E.~Roache$^{4}$,
I.~Sadeh$^{14}$,
L.~Saha$^{4}$,
H.~Salzmann$^{8}$,
M.~Santander$^{22}$,
G.~H.~Sembroski$^{23}$,
B.~Shen$^{12}$,
M.~Splettstoesser$^{8}$,
A.~K.~Talluri$^{9}$,
S.~Tandon$^{19}$,
J.~V.~Tucci$^{24}$,
J.~Valverde$^{25,13}$,
V.~V.~Vassiliev$^{5}$,
D.~A.~Williams$^{8}$,
S.~L.~Wong$^{17}$,
T.~Yoshikoshi$^{26}$\\

\vspace{0.3cm}
\scriptsize
\noindent
$^{1}${Department of Physics and Astronomy, DePauw University, Greencastle, IN 46135-0037, USA}

\noindent
$^{2}${Department of Physics, Temple University, Philadelphia, PA 19122, USA}

\noindent
$^{3}${Department of Physics and Astronomy, University of Utah, Salt Lake City, UT 84112, USA}

\noindent
$^{4}${Center for Astrophysics $|$ Harvard \& Smithsonian, Cambridge, MA 02138, USA}

\noindent
$^{5}${Department of Physics and Astronomy, University of California, Los Angeles, CA 90095, USA}

\noindent
$^{6}${Physics Department, California Polytechnic State University, San Luis Obispo, CA 94307, USA}

\noindent
$^{7}${Department of Physics, Washington University, St. Louis, MO 63130, USA}

\noindent
$^{8}${Santa Cruz Institute for Particle Physics and Department of Physics, University of California, Santa Cruz, CA 95064, USA}

\noindent
$^{9}${School of Physics and Astronomy, University of Minnesota, Minneapolis, MN 55455, USA}

\noindent
$^{10}${Department of Physics and Astronomy, Dartmouth College, 6127 Wilder Laboratory, Hanover, NH 03755 USA}

\noindent
$^{11}${Department of Physics and Astronomy and the Bartol Research Institute, University of Delaware, Newark, DE 19716, USA}

\noindent
$^{12}${Department of Physics, University of Maryland, College Park, MD, USA }

\noindent
$^{13}${NASA GSFC, Greenbelt, MD 20771, USA}

\noindent
$^{14}${DESY, Platanenallee 6, 15738 Zeuthen, Germany}

\noindent
$^{15}${Fakult\"at f\"ur Physik \& Astronomie, Ruhr-Universit\"at Bochum, D-44780 Bochum, Germany}

\noindent
$^{16}${School of Natural Sciences, University of Galway, University Road, Galway, H91 TK33, Ireland}

\noindent
$^{17}${Physics Department, McGill University, Montreal, QC H3A 2T8, Canada}

\noindent
$^{18}${School of Physics, University College Dublin, Belfield, Dublin 4, Ireland}

\noindent
$^{19}${Department of Physics and Astronomy, Barnard College, Columbia University, NY 10027, USA}

\noindent
$^{20}${Institute of Physics and Astronomy, University of Potsdam, 14476 Potsdam-Golm, Germany}

\noindent
$^{21}${Department of Physical Sciences, Munster Technological University, Bishopstown, Cork, T12 P928, Ireland}

\noindent
$^{22}${Department of Physics and Astronomy, University of Alabama, Tuscaloosa, AL 35487, USA}

\noindent
$^{23}${Department of Physics and Astronomy, Purdue University, West Lafayette, IN 47907, USA}

\noindent
$^{24}${Department of Physics, Indiana University Indianapolis, Indianapolis, Indiana 46202, USA}

\noindent
$^{25}${Department of Physics, University of Maryland, Baltimore County, Baltimore MD 21250, USA}

\noindent
$^{26}${Institute for Cosmic Ray Research, University of Tokyo, 5-1-5, Kashiwa-no-ha, Kashiwa, Chiba 277-8582, Japan}
\vspace{1cm}
\section*{Full Author List: \ HAWC Collaboration}
\scriptsize
\noindent
%
\noindent

R. Alfaro$^{1}$,
C. Alvarez$^{2}$,
A. Andrés$^{3}$,
E. Anita-Rangel$^{3}$,
M. Araya$^{4}$,
J.C. Arteaga-Velázquez$^{5}$,
D. Avila Rojas$^{3}$,
H.A. Ayala Solares$^{6}$,
R. Babu$^{7}$,
P. Bangale$^{8}$,
E. Belmont-Moreno$^{1}$,
A. Bernal$^{3}$,
K.S. Caballero-Mora$^{2}$,
T. Capistrán$^{9}$,
A. Carramiñana$^{10}$,
F. Carreón$^{3}$,
S. Casanova$^{11}$,
S. Coutiño de León$^{12}$,
E. De la Fuente$^{13}$,
D. Depaoli$^{14}$,
P. Desiati$^{12}$,
N. Di Lalla$^{15}$,
R. Diaz Hernandez$^{10}$,
B.L. Dingus$^{16}$,
M.A. DuVernois$^{12}$,
J.C. Díaz-Vélez$^{12}$,
K. Engel$^{17}$,
T. Ergin$^{7}$,
C. Espinoza$^{1}$,
K. Fang$^{12}$,
N. Fraija$^{3}$,
S. Fraija$^{3}$,
J.A. García-González$^{18}$,
F. Garfias$^{3}$,
N. Ghosh$^{19}$,
A. Gonzalez Muñoz$^{1}$,
M.M. González$^{3}$,
J.A. Goodman$^{17}$,
S. Groetsch$^{19}$,
J. Gyeong$^{20}$,
J.P. Harding$^{16}$,
S. Hernández-Cadena$^{21}$,
I. Herzog$^{7}$,
D. Huang$^{17}$,
P. Hüntemeyer$^{19}$,
A. Iriarte$^{3}$,
S. Kaufmann$^{22}$,
D. Kieda$^{23}$,
K. Leavitt$^{19}$,
H. León Vargas$^{1}$,
J.T. Linnemann$^{7}$,
A.L. Longinotti$^{3}$,
G. Luis-Raya$^{22}$,
K. Malone$^{16}$,
O. Martinez$^{24}$,
J. Martínez-Castro$^{25}$,
H. Martínez-Huerta$^{30}$,
J.A. Matthews$^{26}$,
P. Miranda-Romagnoli$^{27}$,
P.E. Mirón-Enriquez$^{3}$,
J.A. Montes$^{3}$,
J.A. Morales-Soto$^{5}$,
M. Mostafá$^{8}$,
M. Najafi$^{19}$,
L. Nellen$^{28}$,
M.U. Nisa$^{7}$,
N. Omodei$^{15}$,
E. Ponce$^{24}$,
Y. Pérez Araujo$^{1}$,
E.G. Pérez-Pérez$^{22}$,
Q. Remy$^{14}$,
C.D. Rho$^{20}$,
D. Rosa-González$^{10}$,
M. Roth$^{16}$,
H. Salazar$^{24}$,
D. Salazar-Gallegos$^{7}$,
A. Sandoval$^{1}$,
M. Schneider$^{1}$,
G. Schwefer$^{14}$,
J. Serna-Franco$^{1}$,
A.J. Smith$^{17}$
Y. Son$^{29}$,
R.W. Springer$^{23}$,
O. Tibolla$^{22}$,
K. Tollefson$^{7}$,
I. Torres$^{10}$,
R. Torres-Escobedo$^{21}$,
R. Turner$^{19}$,
E. Varela$^{24}$,
L. Villaseñor$^{24}$,
X. Wang$^{19}$,
Z. Wang$^{17}$,
I.J. Watson$^{29}$,
H. Wu$^{12}$,
S. Yu$^{6}$,
S. Yun-Cárcamo$^{17}$,
H. Zhou$^{21}$

\vskip2cm
\noindent
$^{1}$Instituto de F\'{i}sica, Universidad Nacional Autónoma de México, Ciudad de Mexico, Mexico

\noindent
$^{2}$Universidad Autónoma de Chiapas, Tuxtla Gutiérrez, Chiapas, México

\noindent
$^{3}$Instituto de Astronom\'{i}a, Universidad Nacional Autónoma de México, Ciudad de Mexico, Mexico

\noindent
$^{4}$Universidad de Costa Rica, San José 2060, Costa Rica

\noindent
$^{5}$Universidad Michoacana de San Nicolás de Hidalgo, Morelia, Mexico

\noindent
$^{6}$Department of Physics, Pennsylvania State University, University Park, PA, USA

\noindent
$^{7}$Department of Physics and Astronomy, Michigan State University, East Lansing, MI, USA

\noindent
$^{8}$Temple University, Department of Physics, 1925 N. 12th Street, Philadelphia, PA 19122, USA

\noindent
$^{9}$Universita degli Studi di Torino, I-10125 Torino, Italy

\noindent
$^{10}$Instituto Nacional de Astrof\'{i}sica, Óptica y Electrónica, Puebla, Mexico

\noindent
$^{11}$Institute of Nuclear Physics Polish Academy of Sciences, PL-31342 11, Krakow, Poland

\noindent
$^{12}$Dept. of Physics and Wisconsin IceCube Particle Astrophysics Center, University of Wisconsin{\textemdash}Madison, Madison, WI, USA

\noindent
$^{13}$Departamento de F\'{i}sica, Centro Universitario de Ciencias Exactase Ingenierias, Universidad de Guadalajara, Guadalajara, Mexico

\noindent
$^{14}$Max-Planck Institute for Nuclear Physics, 69117 Heidelberg, Germany

\noindent
$^{15}$Department of Physics, Stanford University: Stanford, CA 94305–4060, USA

\noindent
$^{16}$Los Alamos National Laboratory, Los Alamos, NM, USA

\noindent
$^{17}$Department of Physics, University of Maryland, College Park, MD, USA

\noindent
$^{18}$Tecnologico de Monterrey, Escuela de Ingenier\'{i}a y Ciencias, Ave. Eugenio Garza Sada 2501, Monterrey, N.L., Mexico, 64849

\noindent
$^{19}$Department of Physics, Michigan Technological University, Houghton, MI, USA

\noindent
$^{20}$Department of Physics, Sungkyunkwan University, Suwon 16419, South Korea

\noindent
$^{21}$Tsung-Dao Lee Institute \& School of Physics and Astronomy, Shanghai Jiao Tong University, 800 Dongchuan Rd, Shanghai, SH 200240, China

\noindent
$^{22}$Universidad Politecnica de Pachuca, Pachuca, Hgo, Mexico

\noindent
$^{23}$Department of Physics and Astronomy, University of Utah, Salt Lake City, UT, USA

\noindent
$^{24}$Facultad de Ciencias F\'{i}sico Matemáticas, Benemérita Universidad Autónoma de Puebla, Puebla, Mexico

\noindent
$^{25}$Centro de Investigaci\'on en Computaci\'on, Instituto Polit\'ecnico Nacional, M\'exico City, M\'exico

\noindent
$^{26}$Dept of Physics and Astronomy, University of New Mexico, Albuquerque, NM, USA

\noindent
$^{27}$Universidad Autónoma del Estado de Hidalgo, Pachuca, Mexico

\noindent
$^{28}$Instituto de Ciencias Nucleares, Universidad Nacional Autónoma de Mexico, Ciudad de Mexico, Mexico

\noindent
$^{29}$University of Seoul, Seoul, Rep. of Korea

\noindent
$^{30}$Departamento de Física y Matemáticas, Universidad de Monterrey, Av.~Morones Prieto 4500, 66238, San Pedro Garza Garc\'ia NL, M\'exico

\end{document}